\documentclass[nofootinbib,notitlepage,superscriptaddress,10pt,aps,prd,twocolumn]{revtex4-1}

\usepackage{amsmath,mathtools}
\usepackage{amssymb}
\usepackage{graphicx}
\usepackage{verbatim}
\newcommand{\leri}[1]{\left(#1\right)}
\newcommand{\pbeta}[1]{\prescript{(\beta)}{}{#1}}
\begin{document}

\title{Big-Bounce cosmology in the presence of Immirzi field}
\author{Flavio Bombacigno}
\email{f.bombacigno.pli@gmail.com}
\affiliation{Physics Department, ``Sapienza'' University of Rome, P.le Aldo Moro 5, 00185 (Roma), Italy}
\author{Francesco Cianfrani}
\email{francesco.cianfrani@ift.uni.wroc.pl}
\affiliation{Institute for Theoretical Physics, University of Wroc\l{}aw, Plac\ Maksa Borna
9, Pl--50-204 Wroc\l{}aw, Poland.}
\author{Giovanni Montani}
\email{giovanni.montani@enea.it}
\affiliation{ENEA, FSN-FUSPHY-TSM, R.C. Frascati, Via E. Fermi 45, 00044 Frascati, Italy.\\
Physics Department, ``Sapienza'' University of Rome, P.le Aldo Moro 5, 00185 (Roma), Italy}

\begin{abstract}
The Immirzi parameter is promoted to be a scalar field and the Hamiltonian analysis of the corresponding dynamical system is performed in the presence of gravity. We identified some $SU(2)$ connections, generalizing Ashtekar-Barbero variables, and we rewrite the constraints in terms of them, setting the classical formulation suitable for loop quantization. Then, we consider the reduced system obtained when restricting to a flat isotropic cosmological model. By mimicking loop quantization via an effective semiclassical treatment, we outline how quantum effects are able to tame the initial singularity both in synchronous time and when the Immirzi field is taken as a relational time. 
\end{abstract}

\maketitle

\section{Introduction}
Loop Quantum Gravity \cite{Rovelli:2004tv,Thiemann:2007zz} is 
probably the most valuable attempt to 
canonically quantize the gravitational 
field, essentially in view of its 
well-known successes: the emergence of a 
discrete spectrum of areas and volumes, 
starting from a continuous formulation 
\cite{Rovelli:1994ge} and the rigorous definition of 
a kinematical Hilbert space \cite{Lewandowski:2005jk}, 
allowed by the properties of cylindrical functionals. 

Nonetheless, this proposal is 
affected by non-trivial shortcomings, 
like the difficulties in implementing 
the physical scalar constraint \cite{Thiemann:1997rv}, the 
lack of well-defined classical limit \cite{Thiemann:2002vj}
and, overall the ambiguity of the 
Immirzi parameter choice \cite{Immirzi:1996di}. 

More specifically, different values of 
such a free parameter of the theory 
correspond to deal with non-equivalent 
representations of the quantum picture \cite{Immirzi:1996dr}, 
since they are not connected by a 
unitary transformation \cite{Rovelli:1997na}. 
Over the years many attempts have been 
considered to interpret \cite{Date:2008rb}, 
or to fix \cite{Ghosh:2004wq} the Immirzi parameter (mainly from black hole entropy calculations, even though 
later developments ruled out this possibility \cite{Ghosh:2011fc}). 

Here, we address the point of view to 
treat such a Immirzi variable as a 
real field \cite{Taveras:2008yf,Calcagni:2009xz,Cianfrani:2009sz,Mercuri:2009zt,Lattanzi:2009mg}, whose evolution has 
to be somehow implemented and hopefully 
accounts for the emergence of a given 
constant value. 

In particular in \cite{Mercuri:2009zt} it has been 
demonstrated that the Holst formulation 
for the gravitational action \cite{Holst:1995pc}, 
in the presence of an Immirzi field, 
can be restated as the standard 
Einstein-Hilbert one, plus a 
real massless scalar field action. 

Here, we consider such an issue as our 
starting point and we then re-introduce 
the standard Ashtekar-Barbero-Immirzi 
variables \cite{Ashtekar:1986yd,Barbero:1994ap}, in order to check 
the final structure of the constraints in
Hamiltonian formulation of the theory and to set up the necessary tools for loop quantization.

The aim of this restatement of the Loop 
Quantum Gravity formulation consists 
of checking if the Immirzi field can 
play, in such a scheme, the role of 
a time variable for the gravitational 
field evolution. 


Thus, we consider the implementation of 
the restricted evolutionary theory to 
the quantization of the isotropic 
Robertson-Walker Universe, in order to get 
insight on the nature of the cosmological singularity. As first step toward
such 
an aim, we mimic loop quantization by considering a semi-classical 
polymer approach to the 
considered model. Actually, the obtained 
Hamiltonian possesses non-trivial features, 
making its full quantum treatment almost 
puzzling. 
The present approach to the isotropic 
Universe quantum dynamics is a reliable 
feasibility test on the implementation 
of a quantum Big-Bounce scenario in 
this revised quantum cosmological framework.  The test has been fully
successful 
since we are implementing a well-traced 
Big-Bounce picture, having some peculiarities 
we will discuss in detail below, but 
strongly resembling the one obtained 
by Ashtekar and collaborators in \cite{Ashtekar:2006rx}, 
see also \cite{Ashtekar:2006uz,Ashtekar:2006wn}. 

The here presented issue is really encouraging toward the search for a fully 
quantum implementation of the model 
and the proper construction of a semi-classical limit. 
Finally, we want to stress how the 
considered Immirzi time is promising 
in view of a quantum implementation 
of the so-called Belinski-Khalatnikov-Lifshitz conjecture \cite{bkl1,Belinsky:1982pk}, 
sufficiently near to the singularity, 
when the spatial gradients are 
negligible with respect to  
the system time evolution. In such a 
limit, the Immirzi time should be 
a viable approach and it suggests 
a new general perspective for investigating 
the singularity removal in the 
quantum and semi-classical sector.

\section{Ashtekar-Barbero variables for the Immirzi field}
The action of LQG reads \cite{Holst:1995pc} (in units $c=8\pi G=1$)
\begin{equation}
S_H=\frac{1}{2}\int_{\mathcal{M}}d^4x\;ee^{\mu}_Ie^{\nu}_J\leri{R^{IJ}_{\;\;\;\mu\nu}-\frac{\beta}{2}\epsilon^{IJ}_{\;\;\;KL}R^{KL}_{\;\;\;\mu\nu}},
\label{Holstaction}
\end{equation}
$e_I^{\mu}$ being inverse tetrads of the spacetime manifold and $R^{IJ}_{\;\;\;\mu\nu}$ denotes the curvature of the spin connection $\omega^{IJ}_{\;\;\;\mu}$, {\it i.e.}
\begin{equation}
R^{IJ}_{\;\;\;\mu\nu}=\partial_{[\mu}\omega^{IJ}_{\;\;\;\nu]}+\omega^I_{[\mu K}\omega^{JK}_{\;\;\;\nu]},
\end{equation}
$\beta$ is the inverse of the Immirzi parameter and it multiplies a term (the Holst term) which does not affect the equations of motion. One can also start from Einstein-Hilbert action and show how the Immirzi parameter labels a canonical transformations one can perform on the phase space coordinates. Hence, classically $\beta$ plays no role and one recovers Einstein-Cartan theory. However, the Holst term (or the corresponding canonical transformation) has a nontrivial effect in phase space, since it allows to adopt as variables some $SU(2)$ connections, Ashtekar-Barbero variables \cite{Ashtekar:1986yd,Barbero:1994ap}, and their conjugate momenta $E^a_i$, whose explicit expression reads
\begin{equation}
A^i_a=\frac{1}{\beta}\, K^i_a+\Gamma^i_a\qquad E^a_i=\beta\,\sqrt{q}\,e^a_i\,,
\label{AE}
\end{equation}
$e^a_i$ being inverse triads of the spatial metric $q_{ab}$, while $K^{i}_a$ and $\Gamma^i_a$ are related with the extrinsic and intrinsic curvature of the spatial metric (time and spatial derivatives), respectively.
The constraints becomes 
\begin{align}
&G_i\equiv\mathcal{D}_aE^a_{\;i}=\partial_a E^a_i-\epsilon_{ij}^{\phantom{12}k}A^j_aE^a_k=0\\
&\mathcal{V}_a \equiv F^i_{ab} E^b_i\\
&\mathcal{S}= -\frac{1}{2\sqrt{q}\beta^2}\left\{F^{j}_{\;ab}+\left(1+\frac{1}{\beta^2}\right)\epsilon_{jmn}K^m_{\;a}K^n_{\;b}\right\}\epsilon_{jkl}E^{a}_{\;k}E^{b}_{\;l}\,,
\end{align}
$F^i_{ab}$ being the $SU(2)$ field strength of $A^i_a$. The constraint $G_i$ coincides with the $SU(2)$ Gauss constraint of a Yang-Mills gauge theory, while $\mathcal{V}_a$ and $\mathcal{S}$ are the vector and scalar constraints, which equal, modulo $G_i$, the supermomentum and the superhamiltonian of the metric formulation, respectively.
 
The $SU(2)$ gauge symmetry makes available for quantization some techniques proper of gauge field theories (the use of holonomies and fluxes). Furthermore, on a quantum level, one finds that different values of $\beta$ label inequivalent quantum sectors (for instance the spectrum of the area operator depends on $\beta$ \cite{Rovelli:1994ge}), thus the quantum theory is sensible to the Immirzi parameter. This poses the problem of a classically irrelevant parameter which is a quantum ambiguity. 

In order to tame this un-wanted feature, some approaches have been developed in which the Immirzi parameter is promoted to be a dynamical scalar field \cite{Taveras:2008yf,Calcagni:2009xz,Cianfrani:2009sz,Mercuri:2009zt,Lattanzi:2009mg}. 

In \cite{Calcagni:2009xz} it has been outlined how if a dynamical Immirzi scalar field $\beta=\beta(x)$ is considered in a formulation in which the Holst term is replaced by the Nieh-Yan topological invariant, the system becomes equivalent to Einstein-Cartan theory with a minimally coupled scalar field, {\it i.e.} 
\begin{equation}
S=\frac{1}{2}\int_{\mathcal{M}}d^4x\;e\,e^{\mu}_Ie^{\nu}_J{R}^{IJ}_{\;\;\;\mu\nu}+\frac{3}{4}\int_{\mathcal{M}}d^4x\;e\,\partial_\mu\beta\partial^\mu\beta,
\label{effectiveaction}
\end{equation}
This is the starting point of our analysis. We want to discuss the structure of the phase space in such a theory. Our primary aim is to derive the same kind of $SU(2)$ gauge structure as in Holst formulation, such that the quantization procedure of LQG can be applied also in the presence of the Immirzi field. 
 
In particular, we can get a $SU(2)$ Gauss constraint also for \eqref{effectiveaction}, as soon as the connections and momenta are defined as follows
\begin{align}
&\pbeta{E}^a_{\;i}\equiv\beta(x)\sqrt{q}\,e^a_i\,,
\\
&\pbeta{A}^i_{\;a}\equiv\Gamma^i_{\;a}+\frac{1}{2\beta}\epsilon^{ijk}\,e^b_ke^j_a\partial_b\beta+\frac{1}{\beta}K^i_a\,.
\end{align}
It is worth noting that $\{\pbeta{A}^i_{\;a},\pbeta{E}^a_{\;i}\}$ still form a couple of canonically conjugate variables, {\it i.e.}
\begin{equation}
\{\pbeta{E}^a_i(x),\pbeta{A}^j_b(y)\}=\delta^a_b\delta^j_i\delta(x,y)\,,\label{pbAE}
\end{equation}
and we take them as the coordinates of the gravitational phase space. Other coordinates describe the Immirzi field and we cannot simply take $\{\beta,P\}$, $P$ being the same momentum as that of an ordinary scalar field in metric formulation, since it would have nonvanishing Poisson brackets with $\pbeta{A}^i_{\;a}$. For this reason, we defined the scalar field momentum as follows 
\begin{equation}
\pbeta{P}(x)\equiv P(x)+\frac{1}{\beta}E^a_iK^i_a,
\label{redefP}
\end{equation}
and one can explicitly check that the only nonvanishing Poisson brackets are given by \eqref{pbAE} and 
\begin{equation}
\{\pbeta{P}(x),\beta(y)\}=\delta(x,y)\,.   
\end{equation}

Eventually the vector constraint takes the form
\begin{equation}
\begin{split}
&\mathcal{V}_a\equiv\pbeta{F}^i_{ab}\pbeta{E}^b_i+\pbeta{P}\partial_a\beta,
\end{split}
\label{SupermomentumTot}
\end{equation}
and the scalar constraint reads as
\begin{equation}
\begin{split}
\mathcal{S}\equiv&-\frac{1}{2\beta^2\sqrt{q}}\left\{\prescript{(\beta)}{}{F}^{j}_{\;ab}+\left(1+\frac{1}{\beta^2}\right)\epsilon_{jmn}\,K^m_{\;a}K^n_{\;b}\right\}\times\\&\times\epsilon_{jkl}\prescript{(\beta)}{}{E}^{a}_{\;k}\prescript{(\beta)}{}{E}^{b}_{\;l}\,+\\
&+\frac{3}{4\beta^2\sqrt{q}}\left(1+\frac{1}{\beta^2}\right)\pbeta{E}^a_k\pbeta{E}^b_k\beta_{,a}\beta_{,b}+\\
&-\frac{1}{\beta^3\sqrt{q}}\prescript{(\beta)}{}{E}^a_{\;j}\prescript{(\beta)}{}{E}^b_{\;j}\nabla_a\beta_{,b}\,+\\
&+\frac{1}{3\sqrt{q}}\left(\prescript{(\beta)}{}{P}-\frac{K^{j}_{\;a}\prescript{(\beta)}{}{E}^{a}_{\;j}}{\beta^2}\right)^2\;,
\end{split}
\label{SuperhamiltonianTot}
\end{equation}
where $\pbeta{F}^i_{ab}$ is the $SU(2)$ field strength of $\pbeta{A}^i_a$. 

It is worth noting that the vector constraint \eqref{SupermomentumTot} retains the same form as in Holst formulation in the presence of a minimally coupled scalar field (see \cite{Thiemann:1997rt,Montani:2009gn}). On the contrary, the scalar constraint contains some additional terms. While the theory is classically equivalent to gravity with a minimally coupled scalar field, thus we can perform a change of variables such that the scalar constraint retains the standard form, on a quantum level the choice of variables is crucial and we adopted those suitable for loop quantization. Therefore, while the classical dynamics is equivalent to that of gravity with a minimally coupled scalar field, we expect the quantum dynamics of the Immirzi field to differ significantly. In order to investigate this peculiar dynamics, we consider the symmetry-reduced case of cosmology, in which several simplifications occur.

\section{Minisuperspace model}
Let us consider the homogeneous and isotropic flat Universe described by the FRW line element
\begin{equation}
ds^2=-N^2(t)dt^2+a^2(t)(dx^2+dy^2+dz^2),
\end{equation}
the scale factor $a(t)$ being the only dynamical degree of freedom.
One can choose the triads $e^i_{\;a}=a(t)\delta^i_{\;a}$, such that the pair of conjugate variables $\{\pbeta{E},\pbeta{A}\}$ reduces to
\begin{equation}
\pbeta{E}^a_{\;i}=p\,\delta^a_{\;i}\qquad \pbeta{A}^j_{\;b}=c\,\delta^j_{\;b}\,,
\label{AEcosmological}
\end{equation}
where $\{p,c\}$ are coordinates of the reduced phase space, whose explicit expressions read
\begin{equation}
|p|=|\beta| a^2\qquad c=\frac{\dot{a}}{\beta N} \,,
\end{equation}
and they form a couple of canonical variables with Poisson brackets given by
\begin{equation}
\{p,c\}=\frac{1}{3V_0},
\end{equation}
$V_0$ being the fiducial volume of the considered spacetime region. The scalar field phase space coordinates $\{\beta,\pbeta{P}\}$ are restricted to depend on time only, as well. 

Since all spatial gradients vanish, the vector constraint \eqref{SupermomentumTot} holds identically, while the scalar constraint \eqref{SuperhamiltonianTot} becomes
\begin{equation}
\begin{split}
\mathcal{S}=&-3c^2\sqrt{\frac{|p|}{|\beta|}}(\beta^2-1)-2\sqrt{\frac{|\beta|}{|p|}}\pbeta{P}c+\\
&+\frac{|\beta|^{3/2}}{3|p|^{3/2}}\pbeta{P}^2,
\label{Superhamiltoniancosmoclass}
\end{split}
\end{equation}

It is worth noting that the kinematical structure coincides with that of LQC \cite{Ashtekar:2006uz,Ashtekar:2006wn}, but the scalar constraint differs significantly. In particular, the second term in \eqref{Superhamiltoniancosmoclass} is not present in LQC formulation. 

The equivalence with the classical dynamics of gravity in the presence of a scalar field can be explicitly demonstrated by computing, through Hamilton equations, the Friedman equation, which can be written as
\begin{equation}
\leri{\frac{\dot{a}}{a}}^2=\frac{\pbeta{P}^2}{3p^3}\frac{|\beta|^5}{(1+|\beta|)^2}=\frac{\rho}{3},
\end{equation}
where $\rho$ obeys the continuity equation $\dot{\rho}=-6\rho\frac{\dot{a}}{a}$, which is equivalent to deal with a massless scalar field energy density. Moreover, taking $\beta$ as a clock-like field, one obtains the following dynamics for the scale factor
\begin{equation}
a=a(\beta)=a_0e^{\pm\frac{\beta-\beta_0}{2}}\,,\label{cbt}
\end{equation}
the initial condition being in $a_0=a(\beta_0)$, with the classical singularities in $\beta=\pm\infty$, according to the chosen branch.

We perform a first analysis on the quantum dynamics, by mimicking the quantization procedure adopted in LQC. In particular, we discuss the classical implications of the replacement 
\begin{equation}
c\longrightarrow\frac{\sin \mu c}{\mu},\label{rep}
\end{equation}
in the scalar constraint \eqref{Superhamiltoniancosmoclass}, where the polymer parameter $\mu$ is fixed according to 
\begin{equation}
\mu^2=\frac{\Delta}{|p|^\alpha}\qquad \Delta\equiv 2\sqrt{3}\pi l^2_P\,,
\end{equation}
$\Delta$ being the minimum area gap eigenvalue in LQG, while $\alpha$ is a quantum ambiguity. In particular, in what follows we will consider the most relevant cases in LQC, namely $\alpha=0$ and $\alpha=1$ corresponding to the so-called $\mu_0$ and $\bar\mu$ schemes, respectively.

The motivation for this analysis comes from LQC, where the replacement \eqref{rep} on a classical level is able to capture the main semiclassical nontrivial effect, {\it i.e.} the emergence of the bounce replacing the initial singularity \cite{Singh:2012zc,Diener:2014mia}. 

\begin{figure}
\begin{center}
\includegraphics[height=8.5cm, width=8.5cm,keepaspectratio]{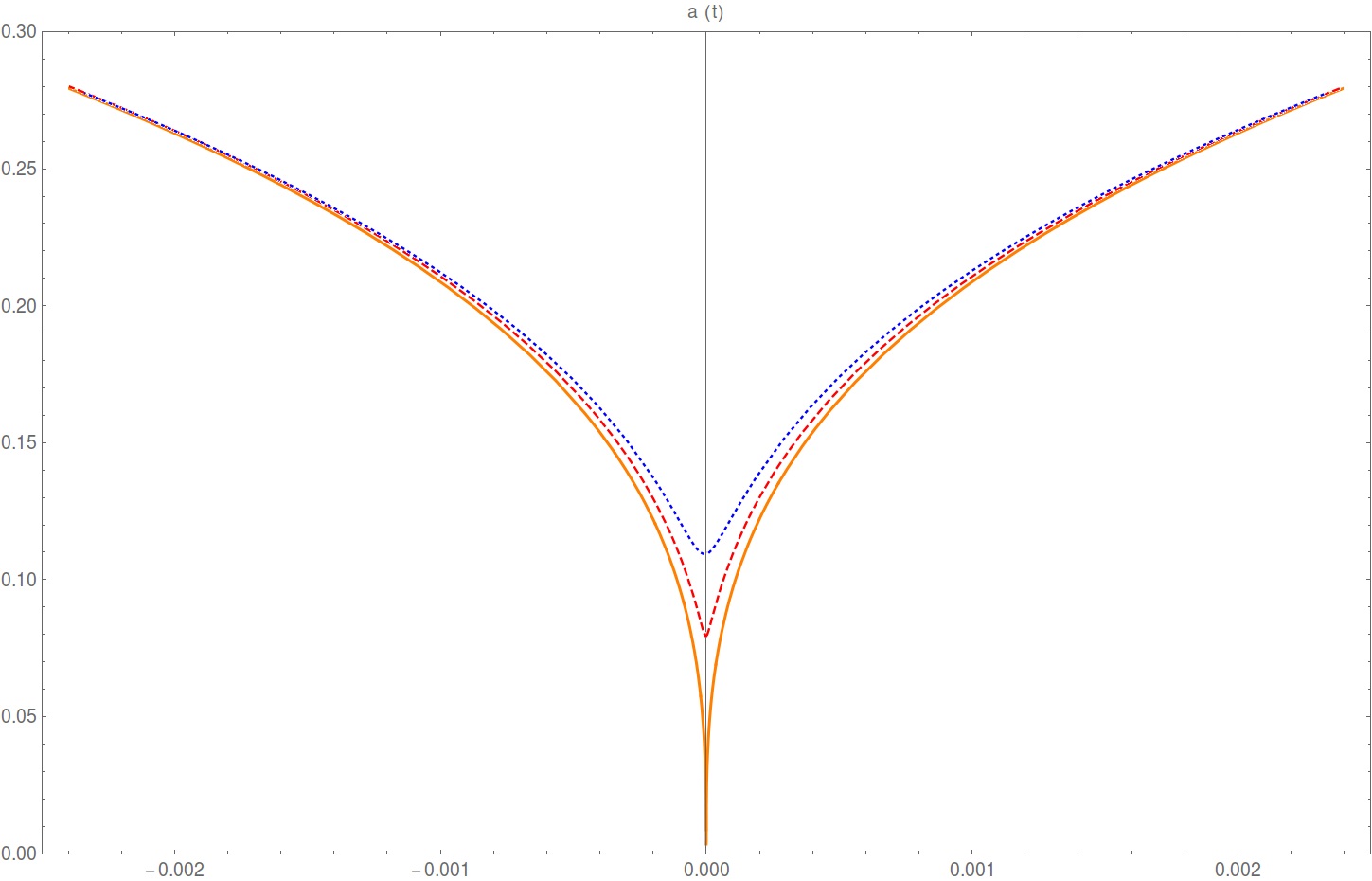}
\caption{Physical scale factor as function of $t$. The solid lines are the two classical branches, reaching the singularity (a=0) for $t=0$. The dotted and dashed lines are the solutions of the polymer cosmological Hamiltonian for $\alpha=0,1$, respectively. Positive and negative branches are plotted and the singularity is smoothly removed matching both the solutions.}
\label{figura3}
\end{center}
\end{figure}

Hence, we consider the classical dynamics generated by the modified scalar constraint 
\begin{equation}
\begin{split}
\mathcal{S}_{sc}=&-3\frac{\sin^2\mu c}{\mu^2}
\sqrt{\frac{|p|}{|\beta|}}(|\beta|^2-1)-2\frac{\sin\mu c}{\mu^2}\sqrt{\frac{|\beta|}{|p|}}\pbeta{P}+\\
&+\frac{|\beta|^{3/2}}{3|p|^{3/2}}\pbeta{P}^2,
\end{split}
\end{equation}
from which the following Friedman equation can be inferred
\begin{equation}
\begin{split}
\leri{\frac{\dot{a}}{a}}^2=\frac{\rho}{3}\leri{\frac{1+|\beta|}{|\beta|}}^2\leri{\sqrt{1-\frac{\rho}{\rho_{C}}}-\frac{|\beta|}{1+|\beta|}}^2,
\end{split}
\label{Friedmanpolymer}
\end{equation}
$\rho_C\equiv\frac{3|\beta|^3}{\mu^2 |p|}$ being the critical density at which the bounce occurs in LQC, and the equations for $\dot{\beta}$ and $\dot{\rho}$ are given respectively by:
\begin{equation}
\dot{\beta}=\frac{2\pbeta{P}}{3p^{3/2}}\frac{|\beta|^{5/2}}{1+|\beta|}\qquad\dot{\rho}=-6\rho\frac{\dot{a}}{a}.
\end{equation}
It is worth noting that the bounce is predicted also from \eqref{Friedmanpolymer}, see figure \ref{figura3}, but at a smaller energy density $^{(\beta)}\rho_{C}$ with respect to LQC:
\begin{equation}
^{(\beta)}\rho_C=\rho_C\frac{|\beta|(|\beta|+2)}{(1+|\beta|)^2}<\rho_C.
\end{equation}
Furthermore the continuity equation for $\rho$ still holds, with the ratio $\frac{\dot{a}}{a}$ now described by \eqref{Friedmanpolymer}.

If we take $\beta$ as a clock-like field, the scale factor behaves as follows
\begin{equation}
a(\beta)=\left[\frac{A}{4 (a_0\beta)^{2+\alpha}}e^{-\frac{2+\alpha}{2}(\beta-\beta_0)}+a_0^{2+\alpha}\,e^{\frac{2+\alpha}{2}(\beta-\beta_0)}\right]^{1/(2+\alpha)}\,,
\end{equation}
remainding that we have to consider just $\alpha=0,1$, where we fixed initial conditions $a(\beta_0)=a_0$ for positive $\beta_0$, while the constant $A$ determine the magnitude of quantum corrections (see the comparison with the $+$ sign classical solution in \eqref{cbt}) and it reads explicitly
\begin{equation}
A= \frac{\Delta \beta_0^2}{9(\beta_0+1)^2}\pbeta{P}^2(\beta_0)\,.
\end{equation}

\begin{figure}
\begin{center}
\includegraphics[height=8.5cm, width=8.5cm, keepaspectratio]{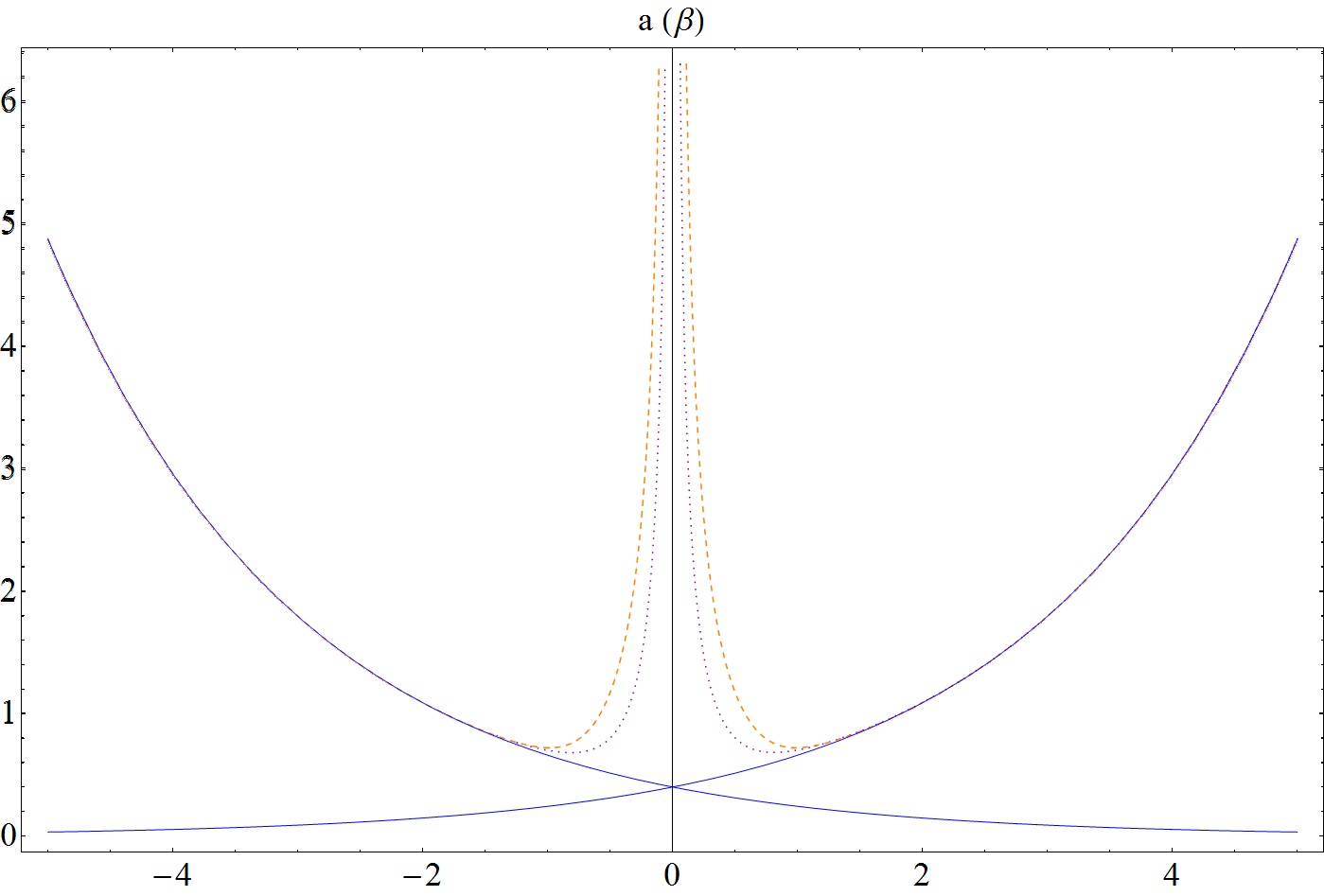}
\caption{Physical scale factor as function of $\beta$: the solid lines are the two classical branches, reaching the singularity (a=0) for $\beta=\pm\infty$.  The dotted and dashed lines are the solutions of the polymer cosmological Hamiltonian for $\alpha=0,1$, respectively. In both cases a divergence in $\beta=0$ arises, while singularity is removed in each solution.}
\label{figura2}
\end{center}
\end{figure}

$a(\beta)$ is plotted in Fig. \ref{figura2}, in which also the solution for negative $\beta_0$ values is drawn, and it outlines how the solution splits in two separate branches for positive and negative values of $\beta$. Each branch has its own bounce and they both diverge in $\beta\rightarrow 0$.

\section{Conclusions}

We performed the Hamiltonian analysis of the model presented in \cite{Calcagni:2009xz}, in which the Immirzi parameter is promoted to be a scalar field. We outlined how it is possible to recover some $SU(2)$ connections, playing the role of Ashtekar-Barbero variables. The corresponding momenta describe the triads of the spatial metric times the squared Immirzi field and emerge as basic variables in loop quantization. Inspired by this achievement we investigate the cosmological implications of the model. We identified the analogous of reduced phase space variables of LQC and we mimic loop quantization by an effective treatment, based on replacing the reduced connection variable by its polymer-like version. We considered two choices of the polymer parameter, corresponding to the so-called $\mu_0$ and $\bar\mu$ schemes in LQC. The results of this analysis shows how the singularity is replaced by a bounce, occurring at lower energy density with respect to the critical energy density in LQC. We also investigated the possibility to take $\beta$ as a relational time. The equation for the scale factor as a function of $\beta$ can be analytically solved and the initial singularity is still replaced by a bounce, but a subtle arises. The scale factor diverges in $\beta=0$, such that one gets two disconnected branches for positive and negative values of the Immirzi field. Further investigations are needed in order to test whether such a divergence is tamed in a full quantum treatment or is a proper feature of the model.

\end{document}